\documentclass[12pt,preprint]{aastex}

\begin{document}
\title{Nature of Cyclical Changes in the Timing Residuals
       from the Pulsar B1642$-$03}
\author{T. V. Shabanova\altaffilmark{}}
\affil{Pushchino Radio Astronomy Observatory, Astro Space Center,
 P. N. Lebedev Physical Institute, Russian Academy of Sciences,
 142290 Pushchino, Russia}
\email{tvsh@prao.ru}

\begin{abstract}
We report an analysis of timing data for the pulsar B1642$-$03
(J1645$-$0317) gathered over the 40-year time span between 1969 and
2008. During this interval, the pulsar experienced eight glitch-like
events with a fractional increase in the rotation frequency
$\Delta\nu/\nu \sim{(0.9-2.6)}\times10^{-9}$. We have revealed two
important relations in the properties of these peculiar glitches.
The first result shows that there is a strong linear correlation
between the amplitude of the glitch and the time interval to the
next glitch with a slope of about $0.0026\times10^{-9}$ Hz
day$^{-1}$. This relation allows us to predict epochs of new
glitches. The second result shows that the amplitude of the
glitches is modulated by a periodic large-scale sawtooth-like
function. As a result of this modulation, the glitch amplitude
varies discretely from glitch to glitch with a step of $1.5\times
10^{-9}$ Hz in the range $(2.4-6.9)\times 10^{-9}$ Hz. The
post-glitch time interval also varies discretely with a step of
$\sim$ 600 days in the range 900--2700 days. An analysis of
the data showed that three modulation schemes with modulation
periods of 43 years, 53 years and 60 years are possible. The
best model is the 60-year modulation scheme including 12 glitches.
We make a conclusion that the nature of the
observed cyclical changes in the timing residuals from PSR
B1642$-$03 is a continuous generation of peculiar glitches whose
amplitudes are modulated by a periodic large-scale sawtooth-like
function. As the modulation function is periodical, the
picture of cyclical timing residuals will be exactly repeated in
each modulation period or every 60 years.
\end{abstract}


\section{Introduction}

The existence of long-term cyclical variations in the spin rate of
PSR B1642$-$03 has been known from the analysis of the first
timing data of the pulsar gathered in 1969--1982
\citep{dow83,dow86}. During this 13-year interval, the timing
residuals of the pulsar exhibited oscillatory but not strictly
periodic variations with an amplitude of about 15 ms and a cycle
duration of about 1000 days \citep{cor85,cor93}. The extension of
the observational interval up to 30 years from 1969 to 1999 has
shown that the timing residuals have a cyclical behavior during
all this interval and the observed variations are characterized by
an amplitude varying from 15 to 80 ms and spacing of maxima
varying from 1000 to 2600 days \citep{sha01}. It has been noticed
that the observed shape of these cyclical residuals does not
depend on the time span of the data analyzed.

The most plausible explanation suggested by the authors
\citep{cor93,sha01} was that the long-term cyclical changes in
residuals could result from free precession. Free precession due
to changes in beam orientation would cause periodic, correlated
changes in both the pulse shape and the first frequency
derivative. Strictly periodic variations in the pulse profile and
pulse arrival times from PSR 1642$-$03 have not been detected.
Nevertheless, robust cyclical behavior in residuals has been
interpreted as evidence for slow precession of the neutron star spin
axis due to a nonspherical shape of the pulsar. Here
we establish that the observed cyclical behavior of the
timing residuals from PSR B1642$-$03 is a result of continuous
generation of peculiar glitches whose amplitudes are modulated
by some periodic large-scale process.

In this paper, the study of the pulsar's rotation behavior is
based on an analysis of the timing data collected for 40 years at
three different observatories. The archival Jet Propulsion Laboratory
(JPL) data include the first timing observations of
the pulsar B1642$-$03, which were carried out at 2388 MHz between
1969 and 1982 using antennas of the Deep Space Network of NASA
\citep{dow83,dow86}. The Jodrell Bank Observatory (JBO) data were
obtained at frequencies of 408, 610, 1400, and 1600 MHz over the
interval 1981--1999 and taken from the earlier paper
\citep{sha01}. The Pushchino Radio Astronomy Observatory (PRAO)
data include the timing data between 1991 and 2008 and a
separate observing session in 1984 September--December.

\section{Observations and Timing Analysis}

Since 1991, the pulsar B1642$-$03 has been observed at 102 and 112
MHz using the Large Phased Array of the Pushchino Observatory,
which is a linearly polarized transit telescope. A 64$\times$20
kHz multi-channel radiometer was used. The resolution of the
recorded signals was either 2.56 or 1.28 ms. The duration of each
session determined by the width of the antenna beam at the source
declination was 3.2 min for the declination of PSR B1642$-$03. The
mean pulse profile in each 20 kHz channel was obtained by
synchronous adding of 500 individual pulses with a predicted
topocentric pulsar period. After dispersion removal, all the
channel profiles were summed to form a mean pulse profile for
the given observing session. The topocentric arrival times of the
pulses for each observing session were calculated by
cross-correlating the mean pulse profile with a standard low-noise
template.

The topocentric arrival times collected at JBO, PRAO, and the
geocentric arrival times obtained from the archival JPL timing
data were all referred to as the barycenter of the solar system at
infinite frequency using the program TEMPO \footnote
{http://www.atnf.csiro.au/research/pulsar/tempo} and the JPL DE200
ephemeris. The coordinates of the pulsar that are required for
this reduction were taken from \citet{hob04}, together with a
proper motion equal to zero \citep{sha01}.

In order to analyze the variations in the pulsar rotation,
a second-order polynomial describing the slow down of
the rotational star was fitted to the experimental data.
The rotational pulse phase $\varphi$ at the barycentric arrival
times $t$ was calculated as
\begin{equation}
\varphi(t) = \varphi_{0} + \nu(t-t_{0}) + {\dot\nu}(t-t_{0})^{2}/2,
\end{equation}
where $\varphi_{0}$, $\nu$, and ${\dot\nu}$ are the pulse phase
(measured in cycles), rotation frequency, and first frequency
derivative at some reference time $t_{0}$, respectively.
 The phase residuals,
obtained as differences between the observed phase and the phase
predicted from a timing model, were used for improving the
spin-down parameters of the pulsar. Pulsar parameters $\nu$ and
$\dot\nu$ measured over the 40-year time span of observations are
given in Section 3. Residuals derived with these new
parameters were used to study variations in the pulsar rotation.

Glitches observed in the rotation frequency of the pulsar
B1642$-$03 are peculiar because their properties differ from those
of normal glitches. Usually normal glitches occur as sudden jumps
in the pulsar's rotation frequency, followed by a post-glitch
relaxation representing a sum of the decaying $\Delta\nu_{d}$ and
the permanent $\Delta\nu_{p}$ components \citep{she96}. The large
glitches show significant exponential decay which is estimated by
the parameter $Q={\Delta\nu_{d}}/{(\Delta\nu_{d}+\Delta\nu_{p})}$.
Normal glitches are characterized by short rise times of less than
one day. In contrast, the pulsar B1642$-$03 exhibits small glitches
that have long rise times of about 400--500 days and show
significant exponential relaxation ($Q\sim0.9$) after a glitch.

\section{Results}

\begin{figure}
\epsscale{.80}
\plotone{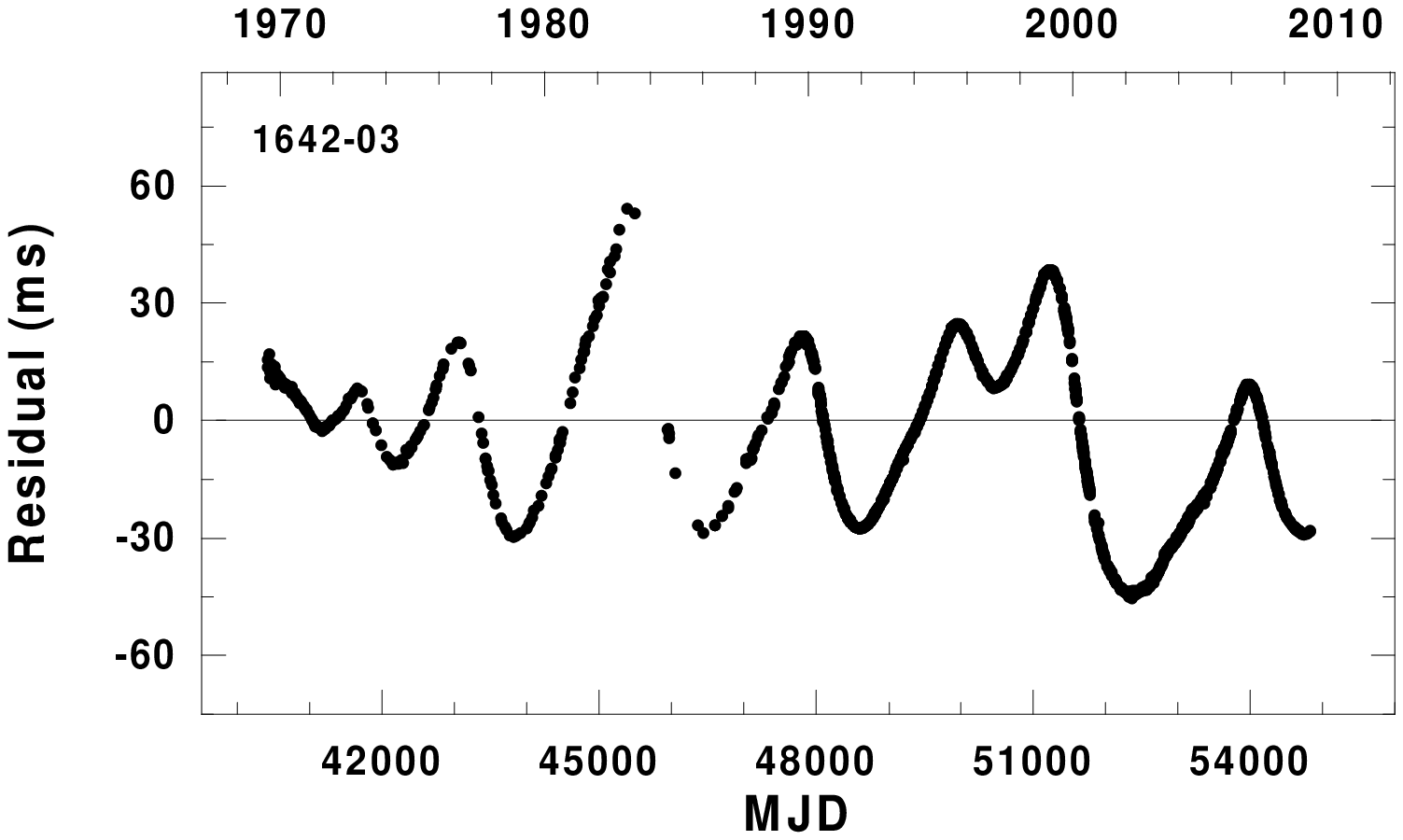}
 \caption{Timing residuals from the pulsar B1642$-$03 over
     the 40-year time span from 1969 to 2008. The residuals,
     derived as the observed times minus the predicted ones,
     are shown after the best fit for $\nu$ and $\dot\nu$
     for all the pulse arrival times. The position was fixed in
     the fitting procedure. A one-year gap between 1983
     and 1984 is seen in the central part of the residual
     curve.\label{resid}}
\end{figure}

The timing residuals after subtraction of the best-fit spin-down
model are presented in Figure~\ref{resid} over the period from
1969 to 2008. It is seen that timing residuals show clear
cyclical behavior over all the 40-year interval. The mean rotation
parameters of the pulsar are very stable and are derived with high
accuracy over the fit interval MJD 40414--54825:
$\nu=2.579388686097(13)$ Hz, $\dot\nu=-11.84578(4)\times
10^{-15}\,s^{-2}$ at the epoch MJD 40414.1297. The measured value
of the second derivative, $\ddot{\nu}\approx
2\times10^{-27}\,s^{-3}$, is mainly determined by an asymmetry of
the residual curve with respect to the $X$-axis over the time span
analyzed. So, a corresponding braking index,
$n={\nu\ddot{\nu}}/{{\dot\nu}^{2}}\sim 30$, may not be related to
the secular slowdown of the pulsar's rotation.

The central part of the residual curve has a one-year gap between
1983 July and 1984 August. The trend of the curve is well traced
on this interval, but in order to study variations in the rotation
frequency of the pulsar in more detail, this gap needs to be
removed. The recovery of the residual curve was based on the
method of prediction of the expected pulse arrival times for particular
epochs. For the interval, which corresponded to the descending
slope of the residual curve between 1983 June and 1986 January,
the values of ${\nu},\,{\dot\nu}$ were determined by the fitting
the timing model. Using the ${\nu},\,{\dot\nu}$ obtained and the
first point of the indicated interval as a reference point, the
expected pulse arrival times were predicted for the epoches spaced by 30
days within the indicated interval. The timing model fitted to all the
points of this interval showed that the timing residuals,
corresponding to the predicted pulse arrival times, coincided
within 1 ms with the residuals, corresponding to the experimental
points from the observing session in 1984 September--December and
the first points of the 1986 data set.

\subsection{The Rotation Behavior of the Pulsar During
            the Period from 1969 to 2008}

\begin{figure}
\epsscale{.80}
\plotone{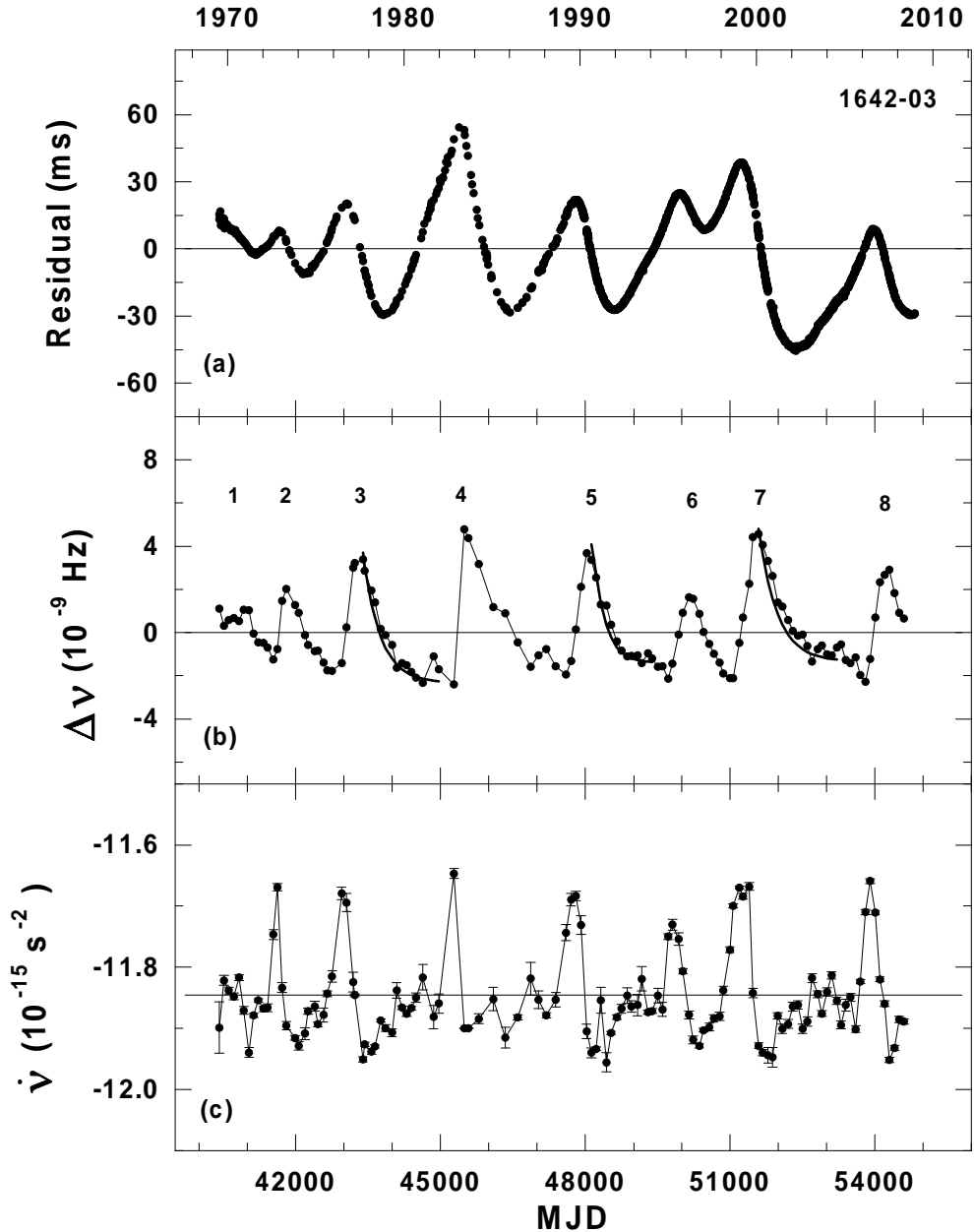}
 \caption{Timing behavior of PSR B1642$-$03 between 1969 and 2008.
     (a) Residuals are the same as in Figure~\ref{resid}, but the gap
     in the data between 1983 and 1986 was removed. It has been filled
     in with points, corresponding to the expected pulse arrival times,
     predicted by the timing model 1983--1986 for the epochs
     spaced by 30 days within this gap.
     (b) The frequency residuals $\Delta\nu$ showing eight peculiar
      glitches. The bold exponential lines fitted to the post-glitch
      points for glitches 3, 5, and 7 indicate a significant exponential
      decay after the glitch with a large value of $Q \sim 0.9$.
     (c) The changes in the frequency first derivative $\dot\nu$
      with time. The peaks of $\Delta\dot{\nu}$ characterize
      the steepness of the front in $\Delta\nu$.
     \label{history}}
\end{figure}

Figure~\ref{history}(a) shows the timing residuals of the pulsar in
which the gap in the data observed was removed.
Figures~\ref{history}(b) and ~\ref{history}(c) show the time
behavior of the frequency residuals $\Delta\nu$ and frequency
derivative $\dot\nu$, respectively. The values of $\nu$ and $\dot{\nu}$
were
calculated from the local fits, performed to arrival time data
over intervals of $\sim$ 200 days that overlapped by 100 days.
The use of the predicted pulse arrival times allowed us to define
rather precisely the epoch and the amplitude of the cycle that
were hidden by a gap in the residual curve.

Figure~\ref{history}(b) shows that the pulsar experienced eight
glitch-like events between 1969 and 2008. These events represent
peculiar glitches because they are characterized by a slow, almost
linear increase in the rotation frequency $\nu$ with a long rise
time of 400--500 days. All these glitches have a small absolute
amplitude observed in the range $\Delta\nu \sim (2.3-6.8)\times
10^{-9}$ Hz. This corresponds to the fractional glitch amplitude
of $\Delta\nu/\nu \sim{(0.9-2.6)}\times10^{-9}$. The largest
glitch size (glitches 4 and 7) is greater than the smallest one
(glitch 1) by a factor of $\sim$ 3.

All the glitches observed exhibit similar post-glitch behavior.
The standard glitch model was not fitted to arrival time data
because of the specific properties of the glitches observed. The
exponential curve was fitted to the frequency residuals
$\Delta\nu$. Figure~\ref{history}(b) shows that the post-glitch
relaxation of glitches 3, 5, and 7 is well described by an
exponentially decaying component with a time constant of $\tau\sim
350-550$ days. Though the exponential curve is not well fitted to
the smaller glitches, it is possible to suppose that all the
glitches observed show a significant exponential decay with a
large value of $Q\sim0.9$.

It is clearly seen from Figure~\ref{history}(c) that the mean value
of the frequency derivative $\dot\nu$, marked on the plot by the
horizontal line, is rather stable over 40 years of observations.
Here, as in the case of slow glitches \citep{sha07}, the increase
in the rotation frequency $\nu$ during the glitch is accompanied
by the decrease in the frequency derivative $\dot\nu$. As is seen
from this plot, the peaks of $\Delta{\dot\nu}$ across the glitch
have an approximately identical magnitude for all the glitches
equal to $\Delta\dot\nu \approx 0.17\times 10^{-15}\,s^{-2}$. This
makes up $\sim1.4\%$ of the mean value of $ \dot\nu\approx
-11.84\times10^{-15}\,s^{-2}$. The peaks of $\Delta\dot\nu$
characterize the steepness of the front in $\Delta\nu$ which
practically does not depend on the glitch amplitude.

An analysis of the changes in $\Delta\nu$ showed that the rotation
frequency of the pulsar B1642$-$03 undergoes continuous generation
of peculiar glitches. A result of this process is clearly seen in
Figure~\ref{history}(b) -- a decrease in $\Delta\nu$ after
one glitch at once passes into an increase in $\Delta\nu$
for the next glitch.

\subsection{The Relation between the Glitch Amplitude and
            the Post-Glitch Interval}

\begin{figure}
\epsscale{.80}
\plotone{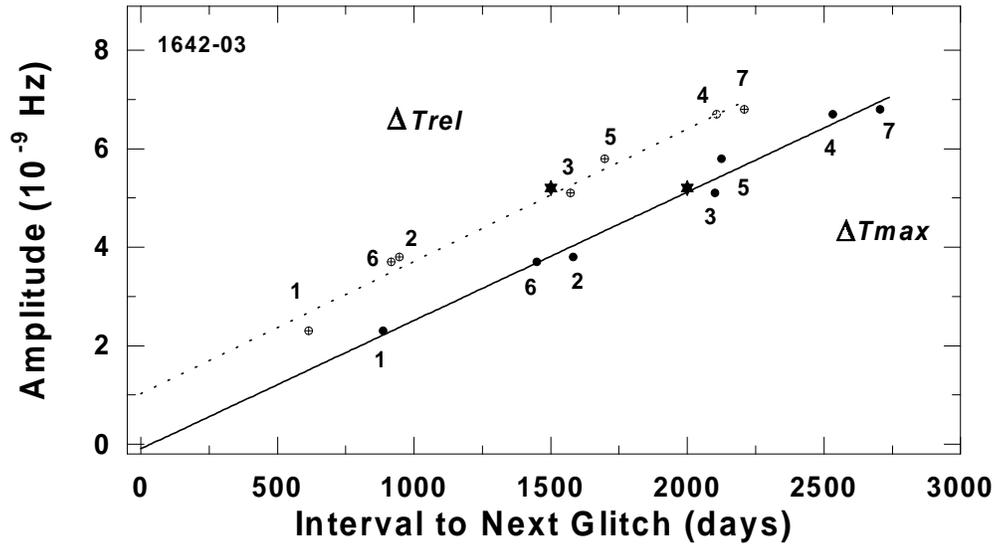}
\caption{Relation between the glitch amplitude and the time
      interval following the glitch. The dotted line corresponds
      to the relation between the glitch amplitude and the relaxation
      time interval $\Delta{T_{rel}}$ (circles around the crosses).
      The solid line corresponds to the relation between the glitch
      amplitude and the time interval to the next glitch
      $\Delta{T_{max}}$ (filled circles). Both the lines have
      a similar slope of about $0.003\times10^{-9}$ Hz day$^{-1}$.
      The solid line $\Delta{T_{max}}$ passes through the origin
      of coordinates. The dotted line $\Delta{T_{rel}}$ indicates
      the existence of a lower bound for the allowed glitch
      amplitudes. The amplitude of glitch 8, marked by
      an asterisk, indicates that the time interval to
      the next glitch is equal to $\sim$ 2000 days.
      \label{relation}}
\end{figure}

Figure~\ref{relation} shows that there is a strong linear
correlation between the glitch amplitude and the time interval
following the glitch. The two curves indicate that the larger is
the glitch amplitude $\Delta{\nu_{g}}$, the larger is the
relaxation time interval after the glitch $\Delta{T_{rel}}$ and
the larger is the time interval to the next glitch
$\Delta{T_{max}}$. We found no evidence of a correlation between
the glitch amplitude and the time interval preceding the glitch.

A linear model was fitted to seven experimental points because
the parameters of the last, eighth glitch are not yet known
completely. The linear relation obtained between the glitch
amplitude $\Delta{\nu_{g}}$ and the post-glitch time intervals is
described by the expressions:
\begin{equation}
\Delta{\nu_{g}} = 0.00261(17)\times \Delta{T_{max}}-0.09(0.32),
\end{equation}
\begin{equation}
\Delta{\nu_{g}} = 0.00269(17)\times \Delta{T_{rel}}+1.02(0.26),
\end{equation}
where uncertainties in the parameters are in parentheses and
refer to the last digits.

Both the fitted straight lines have a similar slope of about
$0.0026(2)\times10^{-9}$ Hz day$^{-1}$ and are spaced
on the $X$-axis by 400--500 days, which is a time of a glitch arising.
The solid line $\Delta{T_{max}}$ passes through the
origin of coordinates. The dotted line $\Delta{T_{rel}}$
indicates the existence of a lower bound for allowed amplitudes.
Glitches, having an amplitude less than $1\times 10^{-9}$ Hz,
should not exist because they will show negative relaxation time.

The linear relation obtained between the glitch amplitude
$\Delta{\nu_{g}}$ and the post-glitch interval to the next glitch
$\Delta{T_{max}}$ allows us to predict epochs of new glitches. In
Figure~\ref{relation}, the amplitude of the eighth glitch marked
by the asterisk indicates that the interval to the next glitch
should be $\sim$ 2000 days. Therefore, the next, ninth glitch will
occur around MJD 56300 or in 2013.

The experimental parameters for eight peculiar glitches plotted in
Figure~\ref{history}(b) are given in Table~\ref{one}. The parameters
are shown in the following order: the glitch number; epoch of the
point ${T_{max}}$, which corresponds to the maximum deviation of
$\Delta{\nu_{max}}$; epoch of the point ${T_{min}}$, which
corresponds to the minimum deviation of $\Delta{\nu_{min}}$; the
glitch amplitude
$\Delta{\nu_{g}}=\Delta{\nu_{max}}+|\Delta{\nu_{min}}|$; the time
interval after the glitch $\Delta{T_{rel}}=T_{min}-T_{max}$; and
the time interval to the next glitch $\Delta{T_{max}}$. As the
eighth glitch still proceeds, some of its parameters are predicted
according to relations (2) and (3). These parameters are printed
bold.

\subsection{The Relation between the Glitch Amplitude and
            the Glitch Number}

\begin{figure}
\epsscale{.80}
\plotone{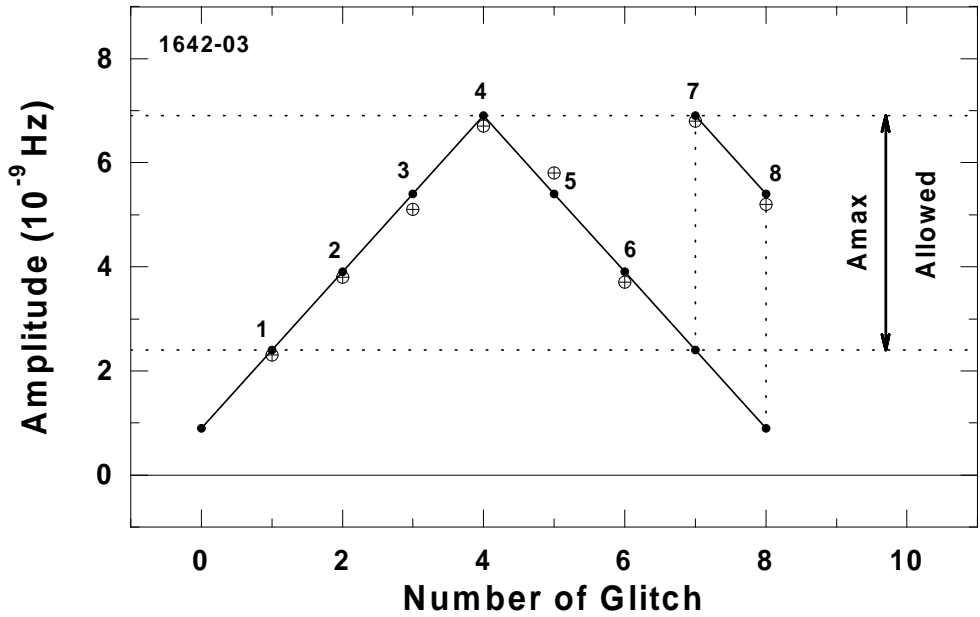}
\caption{Relation between the glitch amplitude and the glitch
    number. The amplitudes of eight glitches observed are marked
    by circles around the crosses. The first six points make up
    the two rectilinear branches which are intersected at point 4.
    Both the branches have an identical slope
    of an opposite sign $a={\pm}1.5\times 10^{-9}$ Hz.
    The predicted amplitudes of eight glitches observed are marked
    by filled circles on these two branches. Points 7 and 8 form
    an additional descending branch, which is parallel to the main
    descending branch. The displacements of points 7 and 8 from
    the predicted values on the main branch are marked by the two
    vertical dotted lines. The two horizontal dotted lines and
    the arrow $A_{max}$ mark the width of the allowed interval
    for the observed glitch amplitudes.
     \label{scheme0}}
\end{figure}

Figure~\ref{scheme0} shows the relation between the glitch
amplitude $\Delta{\nu_{g}}$ and the glitch number in the
sequence of the glitches observed. It is seen that the first
six experimental points, marked by circles around the crosses, make up
two rectilinear branches. The fourth glitch has the maximum
amplitude. After this glitch, the increase of the glitch amplitude
turns into the decrease of the glitch amplitude with the same rate.
The ascending branch (points 1, 2, 3, 4) is
well described by the straight line $y=ax+b$ with the coefficients
$a=1.45(0.04),\,b=0.85(0.11)$, and the descending branch
(points 4, 5, 6) has the coefficients
$a=-1.50(0.35),\,b=8.40(0.75)$.

Figure~\ref{scheme0} clearly shows that the glitch amplitude is
modulated by the sawtooth-like function, both the branches of
which have an identical slope of an opposite sign
$a={\pm}1.5\times 10^{-9}$ Hz. As a result of the modulation, the
amplitude is changed discretely from glitch to glitch with a step
of $1.5 \times 10^{-9}$ Hz as its magnitude depends on the serial
number of the glitch in a given sequence of the glitches. For
further calculations, the ascending branch of the modulation
function will be approximated by a straight line $y=1.5x+0.9$ and
the descending branch by a straight line $y=-1.5x+12.9$, where
$x$ is the glitch number 1,2,3, ...,$n$. The point of intersection of
these two lines has the coordinates, $x_{0}=4.0$ and $y_{0}=6.9$, that
correspond to the observed parameters of the fourth glitch.
Figure~\ref{scheme0} shows that the experimental points well agree
with the points calculated for these two branches.

According to relations (2) and (3), the discrete changes of
the glitch amplitude will cause the
discrete changes of the post-glitch intervals (either increase
or decrease) with a step of $\sim$ 580 days. The discrete
changes of these parameters allow us to estimate more precisely
the lower bound of the allowed glitch amplitudes.
The first glitch in the given sequence of the glitches
has the observed amplitude of $2.3\times10^{-9}$
Hz and is a minimal glitch that can be recorded. A still smaller
glitch should have the amplitude of $\sim 0.8\times10^{-9}$ Hz,
but this glitch cannot exist as it will exhibit a negative
relaxation time interval $\Delta{T_{rel}}$, as it follows from
Figure~\ref{relation}. From here, the allowed interval for the
glitch amplitudes observed is in the range
$(2.4-6.9)\times10^{-9}$ Hz
and has the width equal to ${A_{max}}=4.5\times10^{-9}$ Hz.

As is seen from Figure~\ref{scheme0}, experimental points
7 and 8 produce the second descending branch of the modulation
function that is parallel to the first branch. The derived relation
$y=-1.5x+12.9$ indicates that the predicted amplitudes of these
two glitches should be 2.4 and $0.9\times 10^{-9}$ Hz,
respectively. The comparison of these values with the
values observed, indicated in Table~\ref{one}, shows that the
differences between them make up the same value of about
$\sim4.4\times 10^{-9}$ Hz. This value is very close to the width
of the allowed interval for the glitch amplitudes ${A_{max}}$. We
make a conclusion that the amplitudes observed in glitches 7
and 8 are a result of a forced increase in their initial amplitudes
by the value ${A_{max}}$. Figure~\ref{scheme0} shows that
this unusual phenomenon does not exclude and confirms the existence
of the modulation process.

The phenomenon of a forced increase in the initial amplitudes of
glitches 7 and 8 is also reflected on the timing residuals.
Figure~\ref{history}(a) shows that the pulse arrival times for cycle
7 are earlier as compared with those for cycle 4, though these
glitches have a similar shape in Figure~\ref{history}(b). It looks
as if the pulse arrival times kept the information on partial
identity of the indicated glitches. At the low frequency of 112
MHz, we found no evidence of any changes in the shape or
intensity of the mean pulse profile among cycles 6, 7, and 8.

\subsection{The Modulation Schemes of the Glitch Amplitudes}

A study of the peculiar glitches in the rotation frequency of the
pulsar B1642$-$03 has shown that the amplitudes of these glitches
are modulated by some periodic large-scale sawtooth-like function.
We should define the period and amplitude of this
modulation function. The upper bound of modulation is determined
by the amplitude of the fourth glitch so this glitch is at a cross
point of the ascending and the descending branches of the
modulation function. The glitches with the greater amplitude
should not be observed. The lower bound of the modulation function
is as yet unknown from the observations. An analysis of the data
showed that only three modulation schemes that include an even
quantity of glitches 8, 10, or 12 are possible. A modulation period
cannot include less than eight glitches as this quantity of glitches
is already revealed. Note that the ascending and descending branches
of the modulation function are formed by the predicted magnitudes
of the glitches. The modulation branches, the main and additional,
represent sections of this modulation function. They are located
in the allowed range  and define the observed
magnitudes of the glitches.

\begin{figure}
\epsscale{.80}
\plotone{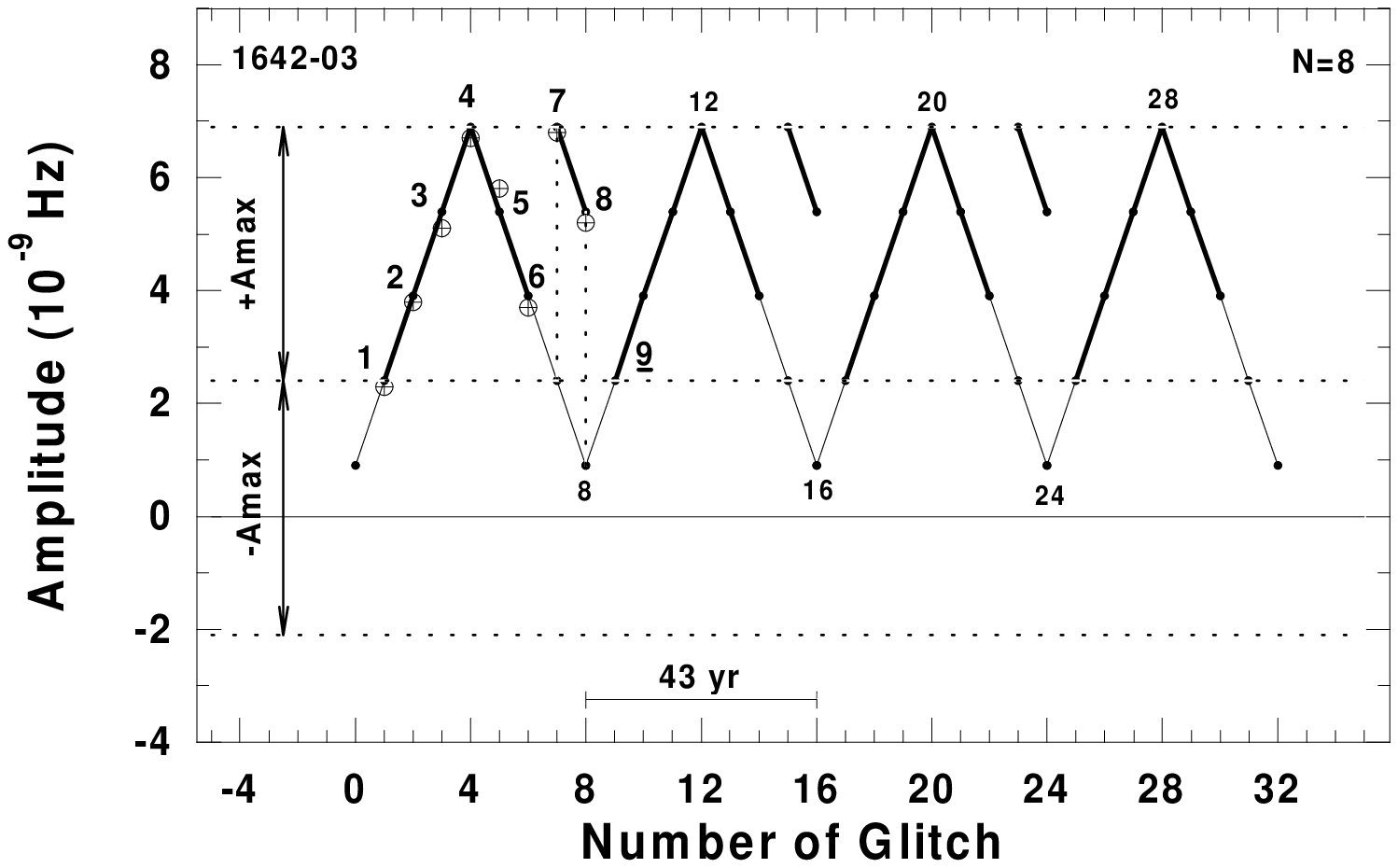}
\caption{Scheme 1 shows four modulation periods, each of which
     includes eight glitches.
     The amplitudes of eight glitches observed are marked by
     circles around the crosses. The three horizontal dotted
     lines and the two arrows $+A_{max}$ and $-A_{max}$ mark
     the width of the allowed interval for the observed
     glitch amplitudes and the width of the forbidden interval,
     respectively. The predicted points on the ascending and
     descending branches are marked by filled circles.
     The two sections of the modulation branches, the main and
     additional, which define the observed amplitudes of
     the glitches, are marked by the bold lines in
     each modulation period. The two vertical dotted lines
     indicate the displacement of points 7 and 8 from
     the predicted values on the main descending
     branch. The full amplitude of the modulation function
     is equal to $\Delta{\nu_{M}}=6\times10^{-9}$ Hz, and
     the modulation period is equal to 43 years.
     Glitch 9 of 2013 will be the first glitch
     on the ascending branch of the second modulation period.
     \label{scheme1}}
\end{figure}

{\bf{Scheme 1}}.
A modulation period includes eight glitches as is shown in
Figure~\ref{scheme1}. It means that each rectilinear branch
will be formed by the amplitudes of the four glitches.
The lower bound of the modulation function will be determined by
the predicted amplitude of the eighth glitch because the eighth
glitch will lie at a cross point of the descending branch of the first
modulation period and the ascending branch of the second modulation
period. This glitch is outside of the allowed interval of the glitch
amplitudes and has the predicted amplitude of $0.9\times10^{-9}$ Hz.
In accordance with this amplitude, the full amplitude of the modulation
function will be equal to $\Delta{\nu_{M}} = 6 \times10^{-9}$ Hz.
In practice, the observed amplitude of the eighth glitch
equals $\sim 5.2 \times10^{-9}$ Hz and is a result of a forced
increase in its initial amplitude by the value ${A}_{max}$.

Table~\ref{two} lists the predicted glitch parameters for three
modulation schemes including 8, 10, and 12 glitches. The predicted
parameters for eight glitches observed are given in the upper part of
this table. The predicted glitch amplitudes were calculated with
the expressions, describing the two branches of the modulation
function: the ascending one as $\Delta{\nu_{g}}=1.5x+0.9$ for
$x=0,1,2,3,4$ and the descending one as
$\Delta{\nu_{g}}=-1.5x+12.9$ for $x=4,5,6,7,8$. The duration of the
post-glitch time intervals $\Delta{T_{rel}}$ and $\Delta{T_{max}}$
were calculated using relations (2) and (3). The epoch of each
glitch $T_{g}$ was calculated by addition of the epoch of the
previous glitch with the corresponding time interval to the next
glitch $\Delta{T_{max}}$. The glitches whose predicted amplitudes
$\Delta{\nu_{g}}$ are in the forbidden range
$-{A_{max}}=[(+2.4)-(-2.1)]\times10^{-9}$ Hz will create an
additional branch of the modulation function. Their expected
observed amplitudes will differ from the predicted one by the
value ${A_{max}}=4.5\times10^{-9}$ Hz. These expected glitch
magnitudes together with the corresponding values of
$\Delta{T_{rel}}$ and $\Delta{T_{max}}$ are printed bold in
parentheses. The predicted parameters for the modulation schemes
including 10 and 12 glitches are presented in the middle and lower
parts of Table~\ref{two}, respectively.

Comparison of the predicted parameters, given in
Table~\ref{two}, and the observed parameters, indicated in
Table~\ref{one}, shows that there is a good agreement among
the parameters $\Delta{\nu_{g}}$, $\Delta{T_{rel}}$, and
$\Delta{T_{max}}$. The predicted glitch epochs $T_{g}$ well
correspond to the observed glitch epochs $T_{max}$ within the
time resolution of $\sim300$ days.

As is seen from Figure~\ref{scheme1}, the ninth glitch will be
the first glitch on the ascending branch of the second modulation
period. The amplitudes of the next glitches will be absolutely
equivalent to the glitch amplitudes of the first modulation
period, as their magnitudes depend only on the serial number of
the glitch in a given modulation period. By our calculations, the
ninth glitch should occur in 2013 (around MJD 56600). If its
observed amplitude is equal to $\Delta{\nu_{g}}=2.4 \times
10^{-9}$ Hz, as indicated in Table~\ref{two}, then the regularity
of the given scheme of the modulation will be confirmed. In this
case, the duration of the modulation period will be 43 years. This
duration is determined by the sum of the post-glitch time
intervals $\Delta{T_{max}}$, indicated in Table~\ref{two}.

\begin{figure}
\epsscale{.80}
\plotone{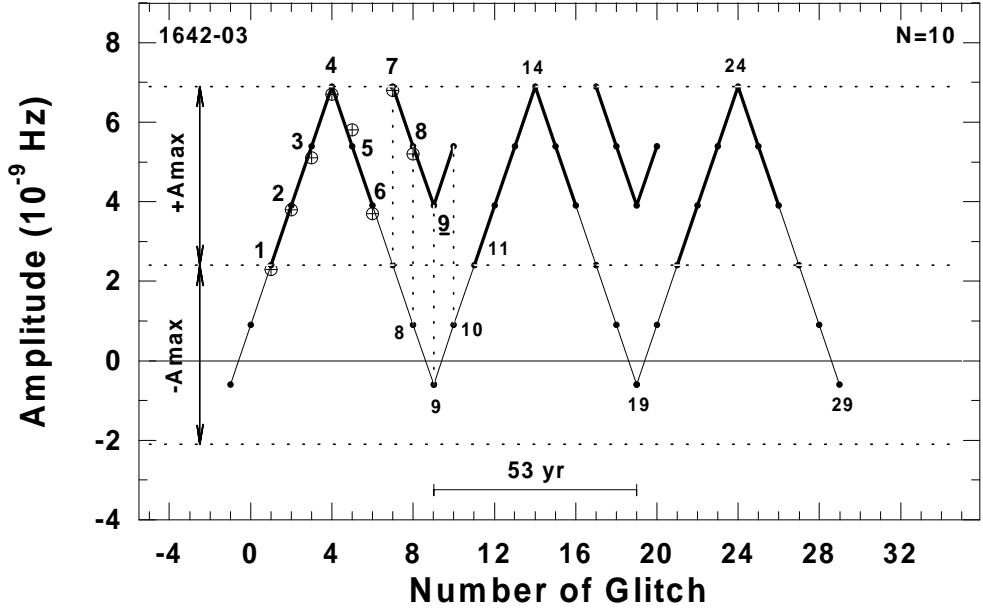}
\caption{Scheme 2 shows three modulation periods, each of which
         includes 10 glitches. The labels are the same as in
         Figure~\ref{scheme1}. The additional modulation branch
         includes four points 7, 8, 9, and 10. The four vertical
         dotted lines indicate the displacement of these points
         from the predicted values on the main branches of
         the modulation. The amplitude of the modulation
         is equal to $\Delta{\nu_{M}}=7.5\times10^{-9}$ Hz, and
         the modulation period is equal to 53 years. Point 9
         marks the amplitude of the next, ninth glitch of 2013.
         Glitch 11 of 2023 will be the first glitch on the ascending
         branch of the second modulation period.
         \label{scheme2}}
\end{figure}

{\bf{Scheme 2}}. A modulation period includes 10 glitches as
is shown in Figure~\ref{scheme2}. It is seen that the predicted
amplitudes of glitches 8, 9, and 10 are outside of the allowed
interval of the glitch amplitudes. Their predicted magnitudes are
given in Table~\ref{two}. The ninth glitch having a negative magnitude
will determine the lower bound of the modulation function. In this
case, the full amplitude of the modulation will be equal to
$\Delta{\nu_{M}} = 7.5 \times10^{-9}$ Hz. The additional branch of
the modulation function will include as many as four glitches.
The expected observed amplitudes of these glitches
will be a result of a forced increase in its initial amplitudes
by the value ${A}_{max}$ and will have the magnitudes that
are printed bold in parentheses in the middle part of Table~\ref{two}.
In this scheme,
the observed amplitude of the next, ninth glitch of 2013 should
be $\Delta{\nu_{g}} = 3.9 \times10^{-9}$ Hz. The duration of the
modulation period will be $\sim$ 53 years.

\begin{figure}
\epsscale{.80}
\plotone{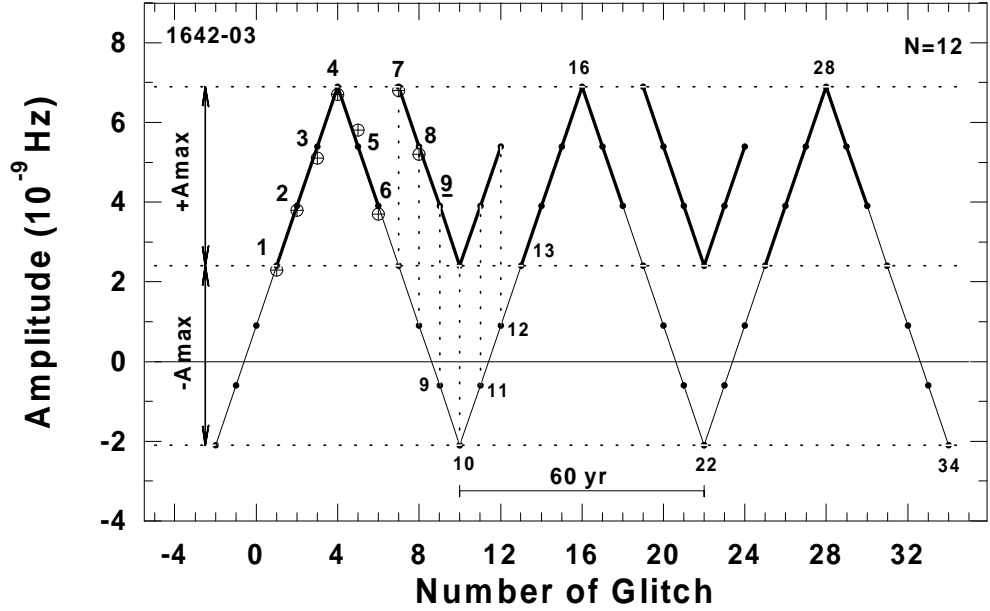}
\caption{Scheme 3 shows three modulation periods, each of which
     includes 12 glitches. The labels are the same as in
     Figure~\ref{scheme1}. The modulation function
     is symmetrical in the range $+A_{max},-A_{max}$. The two
     sections of the modulation branches, the main and additional,
     which are in the allowed range $+A_{max}$, are marked by
     bold lines in each modulation period. The displacements
     of six points of the additional branch from the predicted
     values are marked by six vertical dotted lines.
     It is seen that the additional modulation branch is the mirror
     image of the main modulation branch located in the range
     $+A_{max}$. The full amplitude of the modulation function
     is equal to $\Delta{\nu_{M}}=9\times10^{-9}$ Hz, and
     the modulation period is equal to $\sim$ 60 years.
     Point 9 marks the amplitude of the next, ninth glitch of 2013.
     Glitch 13 of 2030 will be the first glitch on the ascending
     branch of the second modulation period.
         \label{scheme3}}
\end{figure}

{\bf{Scheme 3}}. A modulation period includes 12 glitches as is
shown in Figure~\ref{scheme3}. This modulation scheme looks most
preferable because the modulation function is symmetrical in the
range $+{A_{max}},-{A_{max}}$. The predicted amplitudes of the
first six glitches are in the allowed range
$+{A_{max}}=(6.9-2.4)\times10^{-9}$ Hz and the predicted
amplitudes of the other six glitches are in the forbidden range
$-{A_{max}}=[(+2.4)-(-2.1)]\times10^{-9}$ Hz. According to
relations (2) and (3), the latter glitches cannot exist.
Nevertheless, by analogy with the observed glitches 7 and 8, the
managing process should transfer the amplitudes of these six
glitches to the allowed range by addition of the value $A_{max}$.
Such amplitudes will form the additional branch of the modulation
function. Their magnitudes together with the corresponding values
of $\Delta{T_{rel}}$ and $\Delta{T_{max}}$ are printed bold in
parentheses in the lower part of Table~\ref{two}.
Figure~\ref{scheme3} shows that this additional
modulation branch will be the mirror image of the main modulation
branch. It is clearly seen if we combine point 7 with point 1 and
point 10 with point 4.

In this scheme, the lower bound of the modulation function will be
determined by the predicted amplitude of the 10th glitch
$\Delta\nu_{g}=-2.1\times 10^{-9}$ Hz. Then the full amplitude of
the modulation function will be equal to $\Delta{\nu_{M}} = 9
\times10^{-9}$ Hz, that is, will be equal to the width of the double
interval $2{A_{max}}$. The expected observed amplitude of the
next, ninth glitch of 2013 should be
$\Delta{\nu_{g}}=3.9\times10^{-9}$ Hz and will be the same as in
scheme 2. As is seen from Table~\ref{two}, these two schemes will
differ starting with the 10th glitch that should occur in
$\sim$ 2018. The duration of the modulation period in scheme 3 will
be about 60 years.

Figure~\ref{scheme3} shows that the modulation scheme cannot
include more than 12 glitches. All the predicted glitch amplitudes
should be inside a double interval $2{A_{max}}$, otherwise an
additional modulation branch cannot be formed of the allowed
glitches. Apparently, scheme 3 is the most probable because the
modulation function here is symmetrical and its amplitude is equal
to the full width of the double interval $2{A_{max}}$.

\section{Discussion}

The observed cyclical changes in the timing residuals from PSR
B1642$-$03 is difficult to explain in terms of a free precession
model. Strong evidence for a free precession in the pulsar is
expected to be the detection of the strictly periodic variations
in the timing residuals that should be accompanied by correlated
observable changes in the pulse profile shape
\citep{sha77,nel90,cor93}. A study of the timing behavior of PSR
B1828$-$11 has provided the first evidence of a free precession
in the pulsar \citep{sta00}. The authors have revealed long-term,
strictly periodic, correlated variations in both the pulse arrival
times and the pulse profile and interpreted this phenomenon by
precession of a neutron star spin axis.

In the case of PSR B1642$-$03, no significant changes in the pulse
profile were found within our observations at the frequency of 112
MHz. The pulse profile changes were not detected also in the wide
frequency range 0.1--1.6 GHz \citep{sha01}. The absence of
observable changes in the pulse profile and the presence of
cyclical timing residuals with variable amplitudes and variable
interspaces from three to seven years testify that no significant
precession occurs in this pulsar. As discussed above,
cyclical timing residuals are a result of continuous generation of
peculiar glitches in the pulsar rotation. The finding of the
linear relation between the glitch amplitudes and the post-glitch
intervals indicates that the timing behavior of this pulsar can be
explained well as a glitch phenomenon.

However, note that the timing residuals of the pulsar B1642$-$03
will exhibit strictly periodic changes but with a
very long timescale of about 60 years. As is seen in
Figure~\ref{scheme3}, the glitch amplitudes are modulated by
a periodic large-scale sawtooth-like function. The origin of this
modulation function is as yet unknown. If the pulse profiles
corresponding to the upper and lower parts of this function have
different shapes, we could measure the new pulse shape in the
nearest 20--30 years.

A study of the relation between the glitch amplitude and the time
interval to the next glitch and also to the time interval from the
previous glitch was carried out for pulsars that exhibit multiple
glitches in the rotation frequency \citep{wan00,zou08}. No clear
correlation between the size of the glitch and the corresponding
inter-glitch intervals has been revealed for any of the research
pulsars. The authors supposed that glitches in these pulsars were
due to a local phenomenon, which does not depend on global
stresses.

\begin{figure}
\epsscale{.80}
\plotone{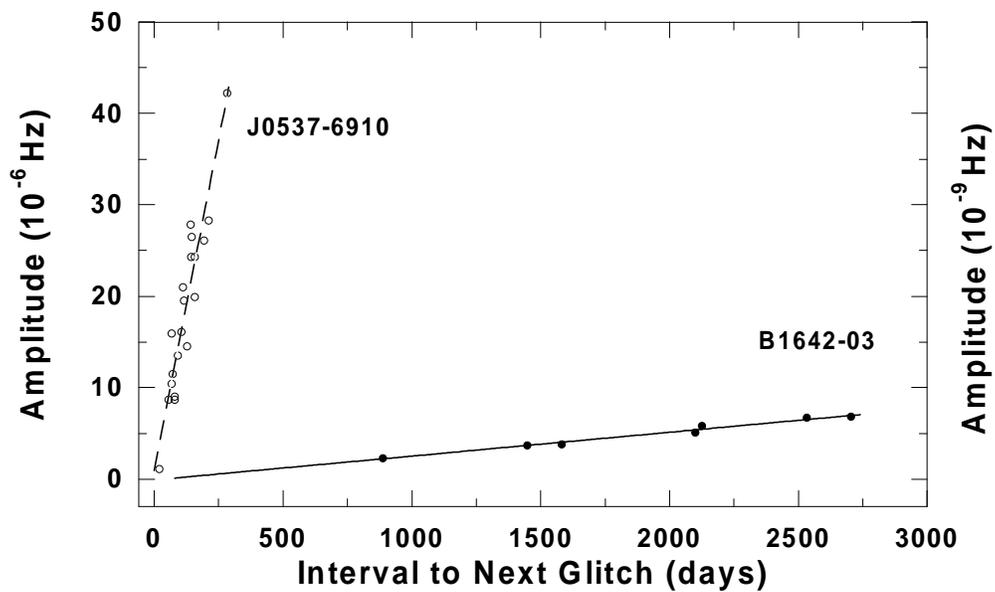}
\caption{Comparison of the two relations showing a correlation
         between the glitch amplitude and the time interval
         to the next glitch for the 16 ms X-ray pulsar
         J0537$-$6910 and for the pulsar B1642$-$03.
         The fitted straight lines have a slope of about
         $0.144\times10^{-6}$ Hz day$^{-1}$ for J0537$-$6910
         and a slope of about $0.003\times10^{-9}$ Hz day$^{-1}$
         for B1642$-$03.
         \label{xray1}}
\end{figure}

The relationship between the size of the glitch and the time
interval to the following glitch was revealed only for the 16 ms
X-ray pulsar J0537$-$6910 \citep{mid06}. During the seven-year period
of observations with the {\it{Rossi X-ray Timing Explorer}}, this pulsar
suffered 21 glitches with a fractional increase in the rotation
frequency ${\Delta\nu}/{\nu}\sim (0.2-6.8)\times10^{-7}$.
Comparison between the glitch parameters for the two pulsars is
presented in Figure~\ref{xray1}. The data observed for the glitches
in J0537$-$6910 were taken from Table~4 of \citet{mid06}.
Note that a characteristic age of the pulsar B1642$-$03 is
$\tau=P/{2\dot{P}} \sim 3.4\times 10^{6}$ years and that of
the pulsar J0537$-$6910 is $\tau\sim 5\times10^{3}$ years.

\begin{figure}
\epsscale{.80}
\plotone{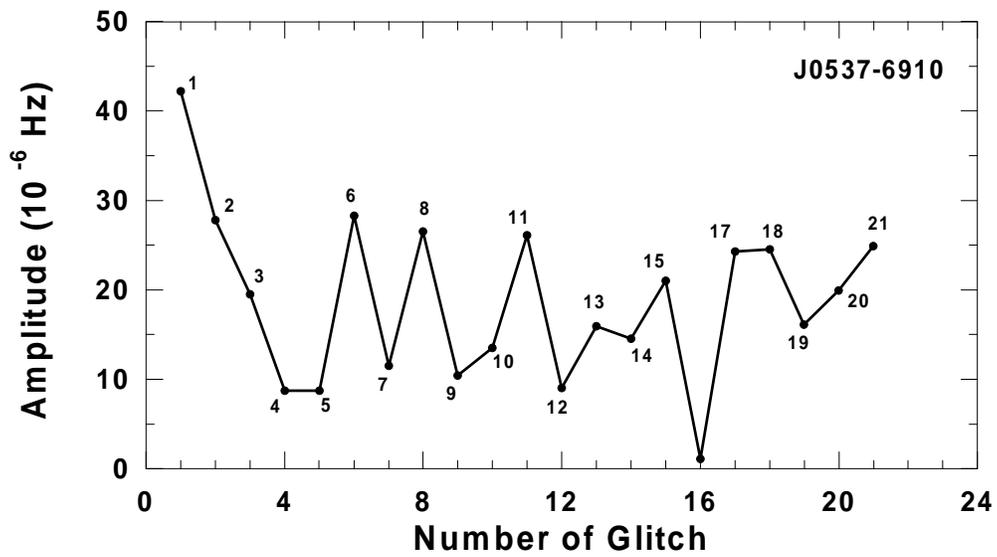}
\caption{Relation between the glitch amplitude and the glitch
         number for J0537$-$6910. Largest glitch 1 was followed
     by a series of 20 glitches with smaller amplitudes.
     The oscillatory behavior of the glitch amplitudes is clearly
     visible between glitches 5 and 12. No obvious modulation
     acting upon the size of the glitches in this pulsar
     is observed.
         \label{xray2}}
\end{figure}

It is seen that the glitch parameters for these two pulsars are at
the different ends of the span of possible magnitudes. The pulsar
J0537$-$6910 shows the largest absolute size of glitches observed
in all pulsars $\Delta\nu \sim (1-42)\times10^{-6}$ Hz. In contrast,
the absolute size of glitches in B1642$-$03 is very small,
approximately four orders of magnitude smaller. In the pulsar
J0537$-$6910, the time intervals between the glitches vary from 20
to 283 days. In contrast, B1642$-$03 presents slow processes,
the time intervals between the glitches nearly 10 times greater
than those seen in J0537$-$6910 and glitches are peculiar, with a
slow, almost linear increase in the rotation frequency during
400--500 days. Nevertheless, both the pulsars exhibit a clear
relation between the glitch size and the time interval to the next
glitch. Figure~\ref{xray1} shows that the fitted straight line has
a slope of about $0.144\times10^{-6}$ Hz day$^{-1}$ for
J0537$-$6910 (or 6.5 days per $\mu$Hz from \citet{mid06}) against
$0.003\times10^{-9}$ Hz day$^{-1}$ for B1642$-$03.
The relation between the glitch amplitude and the glitch number
for J0537$-$6910 is given in Figure~\ref{xray2}. It is seen that
largest glitch 1 was followed by a series of 20 glitches with
the smaller amplitudes. The glitch amplitude has started to oscillate
between glitches 5 and 12. However, there is no indication for
the existence of a modulation process acting upon the size of
the glitches in this pulsar (compare with Figure~\ref{scheme0}).

Glitches are thought to arise from sudden and irregular transfer of
the angular momentum from a more rapidly rotating component of the
superfluid interior to the solid crust of a neutron star. In
terms of vortex pinning models, the origin of glitches can be
explained by the catastrophic unpinning of neutron superfluid
vortices from the lattice of nuclei in the inner crust
\citep{and75,alp84,alp89,alp93,pin85}. This theory provides
a satisfactory explanation for large glitches in pulsars.

The pulsar B1642$-$03 shows small glitches but the properties of
these glitches, such as exponential decay after the glitch and the
existence of a linear relation between the glitch amplitudes and
the relaxation time intervals, well correspond to the
requirements of this theory. In the case of PSR B1642$-$03, it is
necessary to account for the nature of a continuous generation of
peculiar glitches and an origin of a modulation process, which
forces the glitch amplitudes and the inter-glitch intervals to
change with a discrete step. It is also necessary to find out an
interpretation of such an unusual phenomenon of the modulation
process as the transfer of the amplitudes of the glitches, which
are in the forbidden range, to the allowed range by addition of
the value ${A_{max}}$. We make a conclusion that if the pulsar
glitches are due to a variable coupling between the solid crust and
the superfluid interior, then in the case of PSR B1642$-$03 this
variable coupling is provided by the predicted and regular events.

\section{Summary}

An analysis of the timing behavior of PSR B1642$-$03 over the
40-year data span from 1969 to 2008 has shown that the pulsar
rotation frequency is subject to continuous generation of peculiar
glitches whose amplitudes are modulated by some periodic
large-scale process. We pay attention to two aspects of the
phenomenon observed. The first process gives rise to peculiar
glitches having similar properties. These glitches are
characterized by small amplitudes, long rise times of about
400--500 days, and significant exponential decay ($Q \sim 0.9$)
after the glitch. The amplitude of these glitches and the time
interval to the following glitch obey a strong linear relation.

The second process modulates the amplitudes of the peculiar
glitches in such a manner that their magnitudes depend on the serial
number of the glitch in a given modulation period. As a result
of such modulation, the glitch amplitude changes from glitch to
glitch with a discrete step of $1.5\times 10^{-9}$ Hz in the range
$(2.4-6.9)\times 10^{-9}$ Hz. This is accompanied by the
corresponding changes of the time intervals to the following
glitch with a discrete step of $\sim$ 600 days in the range
900--2700 days. We established that the modulation process has a
sawtooth character. The most probable amplitude of this
modulation may be equal to $\Delta{\nu_{M}} = 9 \times10^{-9}$ Hz
and the most probable modulation period may be equal
to $\sim$ 60 years.

Besides, the modulation process gives rise to some additional
modulation branches that are parallel to the main modulation
branches. These branches are composed of the glitches whose
predicted amplitudes should be less than the allowed lower limit
$\sim 2.4\times10^{-9}$ Hz, that is, such glitches should not exist.
Nevertheless, these glitches exist, but they have the amplitudes
that are a result of a forced increase in the predicted
amplitudes by the value ${A_{max}}$.

The nature of cyclical changes in the timing residuals from the
pulsar B1642$-$03 lies in a continuous generation of peculiar
glitches whose amplitudes are modulated by some periodic
large-scale sawtooth-like function. The amplitudes and spacings of
the maxima of the cyclical residuals are a reflection of the
glitch amplitudes and the post-glitch time intervals in the
rotation frequency. The existence of the periodic
sawtooth-like modulation of the glitch amplitudes will cause an
absolutely identical picture of the timing residuals in each
modulation period or every 60 years.

The indicated properties of the peculiar glitches allow us to
predict the epochs and the magnitudes of new glitches in the
rotation frequency of this pulsar, as is shown in
Table~\ref{two}. PSR B1642$-$03 is the first glitching
pulsar that shows that the pulsar glitches can be the
predicted and regular events.

\acknowledgments The author thanks R. D. Dagkesamansky for useful
discussion and comments, the engineering and technical collective
of the PRAO for their aid in
carrying out the many-year observations of this pulsar on the LPA
antenna. The author is grateful to the referee for helpful comments
and suggestions.

%-----------------------------------------------Fig.1
\newpage
\clearpage
%-----------------------------------------------Fig.2
\newpage
\clearpage
%-----------------------------------------------Fig.3
\newpage
\clearpage

%-----------------------------------------------Fig.4
\newpage
\clearpage

%-----------------------------------------------Fig.5
\newpage
\clearpage

%-----------------------------------------------Fig.6
\newpage
\clearpage
%-----------------------------------------------Fig.7
\newpage
\clearpage
%-----------------------------------------------Fig.8
\newpage
\clearpage
%-----------------------------------------------Fig.9
\newpage
\clearpage

%-----------------------------------------------------Table 1
\clearpage
\begin{deluxetable}{lccccccc}
\tablecaption{The Observed Parameters for Eight Peculiar Glitches
        Revealed in PSR B1642$-$03 over the 40-year Interval
	between 1969 and 2008\label{one}}
\tablewidth{0pt}
\tablehead{
\colhead{No.} & \colhead{${T_{max}}$} & \colhead{$\Delta{\nu_{max}}$} &
\colhead{${T_{min}}$} & \colhead{$\Delta{\nu_{min}}$} &
\colhead{$\Delta{\nu_{g}}$} &\colhead{$\Delta{T_{rel}}$} &
\colhead{$\Delta{T_{max}}$}\\
\colhead{} & \colhead{(MJD)} & \colhead{($10^{-9}$ Hz)} & \colhead{(MJD)} &
\colhead{($10^{-9}$ Hz)} & \colhead{($10^{-9}$ Hz)} &
\colhead{(days)} & \colhead{(days)}
}
\startdata
 1   &40920 &1.0 &41534  &-1.3  &2.3  &614  & 886  \\
 2   &41806 &2.0 &42751  &-1.8  &3.8  &945  & 1582 \\
 3   &43388 &3.4 &44959  &-1.7  &5.1  &1571 & 2101 \\
 4   &45489 &4.8 &47596  &-1.9  &6.7  &2107 & 2532 \\
 5   &48021 &3.7 &49719  &-2.1  &5.8  &1698 & 2126 \\
 6   &50147 &1.6 &51062  &-2.1  &3.7  &915  & 1448 \\
 7   &51595 &4.6 &53804  &-2.2  &6.8  &2209 & 2705 \\
 8   &54300 &2.9 &{\bf55850}&{\bf-2.2}&{\bf5.1} &{\bf1520} &{\bf2000} \\
\enddata
\tablecomments{In column order, the table gives the glitch number;
    epoch of the point $T_{max}$, which corresponds to
    the maximum deviation of $\Delta{\nu_{max}}$; epoch of
    the point $T_{min}$, which corresponds to the minimum
    deviation of $\Delta{\nu_{min}}$; the glitch amplitude
    $\Delta{\nu_{g}}= \Delta{\nu_{max}} +|\Delta{\nu_{min}}|$;
    the relaxation time interval after the glitch
    $\Delta{T_{rel}} = T_{min} - T_{max}$; and the time interval
    to the next glitch $\Delta{T_{max}}$. As glitch 8 still
    proceeds, some of its parameters are predicted
    according to relations (2) and (3). They are printed bold.}
\end{deluxetable}

%-----------------------------------------------------Table 2
\clearpage
\begin{deluxetable}{lccccc}
%\tabletypesize{\scriptsize}
\tablecaption{The Predicted Glitch Parameters for Three Modulation
      Schemes Including 8, 10, and 12 glitches
      (see Figures~\ref{scheme1} --~\ref{scheme3}, Respectively)
      \label{two}}
\tablewidth{0pt}
\tablehead{
\colhead{No.} & \colhead{$T_{g}$ (MJD)} &
\colhead{$\Delta{\nu_{g}}$ ($10^{-9}$ Hz)} &
\colhead{$\Delta{T_{rel}}$ (days)} & \colhead{$\Delta{T_{max}}$ (days)} &
\colhead{Date (years)}
}
\startdata
 1 &40920  &2.4         &510  & 960  &1970 \\
 2 &41880  &3.9         &1070 & 1530 &1973 \\
 3 &43410  &5.4         &1620 & 2110 &1977 \\
 4 &45520  &6.9         &2180 & 2690 &1983 \\
 5 &48210  &5.4         &1620 & 2110 &1990 \\
 6 &50320  &3.9         &1070 & 1530 &1996 \\
 7 &51850  &2.4 {\bf (6.9)} &510 {\bf (2180)}& 960 {\bf (2690)}&2000 \\
 8 &54540  &0.9 {\bf (5.4)} &-40 {\bf (1620)}& 380 {\bf (2110)}&2008 \\
 {\bf9}    &56650  &2.4     &510  & 960  {\bf{scheme 1}}&2013\\
\tableline
 8 &54540  &0.9 {\bf (5.4)} &-40 {\bf (1620)}& 380 {\bf (2110)}&2008 \\
 9&56650  &-0.6{\bf (3.9)} &-600 {\bf (1070)}&-196 {\bf (1530)}&2013 \\
 10&58180 &0.9  {\bf (5.4)}&-40 {\bf (1620)} & 380 {\bf (2110)}&2018 \\
 {\bf11}  &60290 &2.4       &510   &960  {\bf{scheme 2}}&2023\\
\tableline
 8 &54540  &0.9 {\bf (5.4)} &-40 {\bf (1620)}& 380 {\bf (2110)}&2008 \\
 9&56650  &-0.6{\bf (3.9)} &-600 {\bf (1070)} &-196 {\bf (1530)}&2013 \\
 10&58180 &-2.1{\bf (2.4)} &-1160 {\bf (510)} &-773 {\bf (960)}&2018 \\
 11&59140 &-0.6{\bf (3.9)} &-600 {\bf (1070)} &-196 {\bf (1530)}&2020 \\
 12&60670 &0.9 {\bf (5.4)} &-40 {\bf (1620)} &380 {\bf (2110)}&2024 \\
 {\bf13}  &62780 &2.4      &510   &960  {\bf{scheme 3}}&2030\\
\enddata
\tablecomments{In column order, the table gives the glitch number, the glitch
      epoch $T_{g}$, the glitch amplitude $\Delta{\nu_{g}}$,
      the relaxation time interval after the glitch $\Delta{T_{rel}}$,
      the time interval to the next glitch $\Delta{T_{max}}$, and
      the glitch date. The amplitudes were
      calculated as $\Delta{\nu_{g}}=1.5x+0.9$ for $x=0,1,2,3,4$
      and as $\Delta{\nu_{g}}=-1.5x+12.9$ for $x=4,5,6, ...,10$.
      The intervals $\Delta{T_{rel}}$ and $\Delta{T_{max}}$ were
      calculated using relations (2) and (3).
      The epoch $T_{g}$ was calculated by addition of the epoch of
      the previous glitch with the corresponding interval
      $\Delta{T_{max}}$. If the predicted $\Delta{\nu_{g}}$ was
      in the forbidden range $-{A_{max}}=[(+2.4)-(-2.1)]\times 10^{-9}$ Hz,
      it was increased by the value $4.5\times10^{-9}$ Hz and
      was printed bold in parentheses together with its corresponding
      values of $\Delta{T_{rel}}$, $\Delta{T_{max}}$.
      The first glitch on the ascending branch of the second
      modulation period was glitch 9 for scheme 1, glitch 11 for
      scheme 2, and glitch 13 for scheme 3.}
\end{deluxetable}

\end{document}